\documentclass[fleqn,usenatbib,useAMS,twocolumn]{aastex63}

\usepackage{epstopdf}
\usepackage{multimedia}
\usepackage{epsfig,graphics,subfigure,psfrag,amsmath,amssymb,amscd}
\usepackage{url, hyperref, fancyhdr}

\usepackage{acronym}
\usepackage{ulem, cancel} 

\newcommand{\D}{\mathrm{d}}

\acrodef{GW}{gravitational wave}
\acrodef{EM}{electromagnetic}
\acrodef{BNS}{binary neutron star}
\acrodef{NSBH}{neutron star-black hole binary}
\acrodef{BBH}{binary black hole}
\acrodef{SBBH}{stellar-mass binary black hole}
\acrodef{ET}{Einstein Telescope}
\acrodef{CE}{Cosmic Explorer}
\acrodef{CI}{confidence interval}
\acrodef{CDF}{cumulative distribution function}
\acrodef{MCMC}{Markov Chain Monte Carlo}
\acrodef{SNR}{signal-to-noise ratio}
\acrodef{AGN}{active galactic nuclei}
\acrodef{CMB}{cosmic microwave background}

\begin{document}

\title{The Dark Side of Using Dark Sirens to Constrain the Hubble-Lema\^itre Constant}


\author{Liang-Gui Zhu}
\thanks{Boya fellow}
\affiliation{Kavli Institute for Astronomy and Astrophysics, Peking University, \\
Beijing 100871,  People's Republic of China. 
\href{Corresponding author.}{xian.chen@pku.edu.cn}}
\affiliation{TianQin Research Center for Gravitational Physics $\&$ School of Physics and Astronomy, \\
 Sun Yat-sen University (Zhuhai Campus), Zhuhai 519082,  People's Republic of China. }

 \author{Xian Chen}
\affiliation{Kavli Institute for Astronomy and Astrophysics, Peking University, \\
Beijing 100871,  People's Republic of China. 
\href{Corresponding author.}{xian.chen@pku.edu.cn}}
\affiliation{Department of Astronomy, School of Physics, Peking University, Beijing 100871,  People's Republic of China. }

\begin{abstract}
Dark sirens, i.e., gravitational-wave (GW) sources without electromagnetic
counterparts, are new probes of the expansion of the Universe.
The efficacy of this method relies on correctly localizing the host galaxies. 
However, recent theoretical studies have shown that astrophysical environments 
could mislead the spatial localization by distorting the GW signals.
It is unclear whether and to what degree the incorrect
spatial localizations of dark sirens would impair the accuracy of the measurement of the cosmological
parameters. To address this issue, we consider the future observations 
of dark sirens using the Cosmic Explorer and the Einstein Telescope, and we
design a Bayesian framework to access the precision of measuring the Hubble-Lema\^itre constant $H_0$. Interestingly, we find that the precision is not compromised when the number of well-localized dark sirens is significantly below $300$, even in the extreme
scenario that all the dark sirens are localized incorrectly.
As the number exceeds $300$, the incorrect spatial localizations start to produce
statistically noticeable effects, such as a slow convergence of the posterior
distribution of $H_0$. 
We propose several tests that can be used in future 
observations to verify the spatial localizations of dark sirens.
Simulations of these tests suggest that incorrect spatial localizations
will dominate a systematic error of $H_0$ if as much as $10\%$
of a sample of $300$ well-localized dark sirens are affected by their environments.
Our results have important implications for the long-term goal of measuring
$H_0$ to a precision of $<1\%$ using dark sirens. 
\end{abstract}

\keywords{Gravitational waves (678), Hubble-Lema\^itre constant (758), Stellar-mass black holes (1611), Sky surveys (1464), Bayesian statistics (1900)}

\section{Introduction}    \label{sec:Introduction}

The detection of \acp{GW} by the Laser Interferometer Gravitational-wave
Observatory (LIGO) and the Virgo detectors \citep{LIGOScientific:2018mvr,
LIGOScientific:2021usb, LIGOScientific:2021djp} opened the possibility of using
\acp{GW} as a new messenger to measure the geometry of the Universe. The idea
was proposed in \citet{Schutz:1986gp} where the author suggested combining the
luminosity distance of a \ac{GW} source and the redshift of its \ac{EM} counterpart to
measure the expansion of the Universe.  Such an idea, more widely known as the
``standard siren'' \citep{Holz:2005df}, was put to the test immediately after the detection of a binary
neutron star merger \citep{LIGOScientific:2017zic, LIGOScientific:2017ync,
Hjorth:2017yza,LIGOScientific:2017adf, LIGOScientific:2018gmd,
LIGOScientific:2019zcs, LIGOScientific:2021aug, DES:2019ccw, DES:2020nay,
Vasylyev:2020hgb, Finke:2021aom}, which by now has achieved a precision of
about $10\%$ \citep{LIGOScientific:2021aug}
for the measurement of the Hubble-Lema\^itre constant,
or better if provided additional information such as observations of jets
\citep{Hotokezaka:2018dfi, Mukherjee:2019qmm}. 
Given the current discrepancy at $4\sigma$
confidence level between the Hubble-Lema\^itre constants measured from the
observations of type-Ia supernovae and the \ac{CMB} \citep{Freedman:2017yms,
Riess:2019qba, DiValentino:2021izs, Perivolaropoulos:2021jda}, it is expected
that standard sirens, as an independent method, could help resolve the tension
\citep{KAGRA:2013rdx, LIGOScientific:2016wof, Reitze:2019iox, Punturo:2010zz,
Chen:2020zoq}.

To qualify as a standard siren, the source should provide observables
from which both the luminosity
distance and the redshift can be obtained. While the luminosity distance is a direct observable from \ac{GW}
signal, the redshift is not always acquirable. In particular, mergers of \acp{BBH}, although 
predominant  among the LIGO/Virgo event, 
are generally not expected to emit detectable \ac{EM} radiation
\citep{Schutz:1986gp} and hence are normally referred to as ``dark sirens''
\citep[or ``dark standard sirens'',][]{DES:2019ccw}.  

The redshifts of dark sirens need to be
obtained by alternative means. One standard method is
marginalizing the redshifts of the galaxies
in the error volume 
pinned down by \ac{GW} detectors
\citep[][]{Schutz:1986gp, MacLeod:2007jd, Petiteau:2011we, DelPozzo:2011vcw,
LIGOScientific:2018gmd, DES:2019ccw, DES:2020nay, LIGOScientific:2019zcs,
LIGOScientific:2021aug}. Such 
a marginalization could introduce large statistical errors
in the resulting cosmological parameters, because there are normally $10^2$ to a
few$\times 10^5$ galaxies in the error volume localized by each dark siren.
However, the overwhelmingly large number of dark sirens
\citep{Maggiore:2019uih, Evans:2021gyd} could compensate for the weakness of
each 
one and thus altogether provide a precise measurement of the
cosmological parameters. As the number of dark sirens increases, the precision
could be comparable with, if not better than, the precision using a more
limited sample of (bright) standard sirens, as recent studies suggest
\citep{DelPozzo:2011vcw, Chen:2017rfc, LIGOScientific:2018gmd, Gray:2019ksv,
Borhanian:2020vyr, Yu:2020vyy, Muttoni:2021veo, Zhu:2021bpp}.

To better estimate the redshifts of dark sirens,
more sophisticated methods have been devised
\citep{LIGOScientific:2018gmd, Gray:2019ksv, Finke:2021aom, Zhu:2021bpp, Gupta:2022fwd}. 
These methods are based on a key assumption, which is that the error volume localized
by \ac{GW} detectors truly contains the host galaxy of a dark siren.
This assumption may not always be valid in light of the recent theoretical discovery
that \ac{GW} signals could have been distorted by the astrophysical environments
in which the waves are produced and propagating \citep[see][for a summary]{Chen:2020iky}. 
For example, 
strong gravitational lensing could make a \ac{GW} source appear more distant \citep{Smith:2017mqu,
Broadhurst:2018saj}. Gravitational redshift  \citep{Chen:2017xbi, Chen:2020iky} or
acceleration of the source around a massive body \citep{Robson:2018svj,
Tamanini:2019usx, Chen:2020iky}, as well as the gas surrounding the source \citep{Chen:2019jde, Chen:2020lpq, Caputo:2020irr}, could bias the measurement of the mass of the source and, in turn,
induce an error in the inference of the luminosity distance.
In addition, the eccentricity of the orbit of a binary 
\citep{Gayathri:2020mra, Gayathri:2020coq} and the  nonstationary noise of
detectors \citep{Edy:2021par, Kumar:2022tto}, if not appropriately account for,
could also result in biased localization of the source.

Therefore, it is possible that unmodeled astrophysical factors could lead us to
an error volume that does not contain the true spatial position of a dark siren. In
this case, a redshift can still be compiled using the galaxies in the incorrect
volume. But the redshift is physically unrelated to the
luminosity distance (incorrectly) inferred from the \ac{GW} signal.  It is
unclear how common such a mismatch is and how seriously it could hamper the
effort of accurately measuring the cosmological parameters using dark sirens.

Here, we try to address the above questions.  The paper is organized as follows.
In Section~\ref{sec:Methodology}, we introduce the Bayesian framework that we
use to infer cosmological parameters from the observations of dark sirens.  In
Section \ref{sec:Simulations}, we conduct mock observations assuming next-generation \ac{GW} detectors.  The results are shown in
Section~\ref{sec:ResultsAnalyses}, where we also test the methods of 
verifying the reliability of the candidate host galaxies of dark sirens. 
In Section~\ref{sec:Discussions}, we discuss the dependence on the fraction of false candidate host galaxies for unbiased cosmological parameter estimation. 
Finally, in Section~\ref{sec:Conclusions}, we summarize the main findings of this work and conclude with the implications of our results.

\section{Methodology}    \label{sec:Methodology}

\subsection{Cosmological model}

We assume that the expansion of the late Universe can be described by a
flat $\Lambda$ cold dark matter (flat-$\Lambda$CDM) cosmological model. The
evolution of the expansion rate with redshift can be expressed
as 
\begin{equation}  \label{eq:H_z} 
H(z) = H_0 \sqrt{\Omega_M(1+z)^3 +
\Omega_{\Lambda}}, 
\end{equation} 
where $H_0 \equiv H(z=0)$ is the Hubble-Lema\^itre constant describing the
current expansion rate of the Universe, $\Omega_M$ is the matter density
normalized by the critical density, and $\Omega_\Lambda \equiv 1- \Omega_M$ is
the fractional density of dark energy. 

Because we are interested in the relationship between 
the luminosity distance $D_L$ of a \ac{GW} source and the redshift $z$ of its host galaxy,
we use
\begin{equation}  \label{eq:DL-z} 
D_L =c(1+z) \int_{0}^{z} \frac{1}{H(z')} \D z' , 
\end{equation} 
to link the two quantities,
where $c$ is the
speed of light.  
In this cosmological model, the set of independent parameters are 
$\vec H \equiv \{H_0, \Omega_M \}$.  

\subsection{Bayesian framework}   \label{sec:Bayesian}

To extract the cosmic expansion information from \ac{GW} and \ac{EM} data, we adopt the Bayesian framework presented in \cite{Chen:2017rfc} \cite[also see][]{Gray:2019ksv, Finke:2021aom, Zhu:2021bpp}.  
This framework uses a data set composed of $N$ independent \ac{GW} events $\mathcal{D}^{\rm GW} \equiv \{ d^{\rm GW}_1, d^{\rm GW}_2, \ldots, d^{\rm GW}_i, \ldots, d^{\rm GW}_N \}$ 
as well as the corresponding \ac{EM} counterparts $\mathcal{D}^{\rm EM} \equiv \{ d^{\rm EM}_1, d^{\rm EM}_2, \ldots, d^{\rm EM}_i, \ldots, d^{\rm EM}_N \}$ to
infer the probability distribution (known as the posterior) of the cosmological parameters. Using the above notations, the posterior can be written as
\begin{align}  \label{eq:posterior} 
p(\vec{H} |\mathcal{D}^{\rm GW}, \mathcal{D}^{\rm EM}, I) \propto p_0(\vec{H}|I) p(\mathcal{D}^{\rm GW},
\mathcal{D}^{\rm EM}|\vec{H}, I)  \nonumber \\ 
\propto p_0(\vec{H}|I) \prod_i p(d^{\rm GW}_i, d^{\rm EM}_i|\vec{H}, I), 
\end{align} 
where $I$ indicates all the relevant additional information. 

The term $p(d^{\rm GW}_i, d^{\rm EM}_i|\vec{H}, I)$ denotes the likelihood of
the data for the $i$th event.  Because the data contain noise, different sources
with different luminosity distances $D_L$, redshifts $z$, sky longitudes
$\alpha$ and latitudes $\delta$ may result in the same data.  Therefore, the
likelihood is a marginalization of all possible $(D_L, z, \alpha, \delta)$,
i.e.,
\begin{align}  
\label{eq:likeli_decompos}
&p(d^{\rm GW}_i, d^{\rm EM}_i| \vec{H}, I)  \nonumber \\
=& ~\frac{\int p(d^{\rm GW}_i, d^{\rm EM}_i, D_L, z, \alpha, \delta | \vec{H}, I) \D D_L  \D z  \D \alpha  \D\delta }{\beta(\vec{H} | I)}. 
\end{align}
Here, $\beta(\vec{H} | I)$ is a normalization factor that accounts for the
selection effects in \ac{GW} \citep{Fishbach:2017zga, LIGOScientific:2018mvr,
Mandel:2018mve} and \ac{EM} observations \citep{Chen:2017rfc}.  To calculate
the probability $p(d^{\rm GW}_i, d^{\rm EM}_i, D_L, z, \alpha, \delta|\vec{H},
I)$, we separate it into
\begin{align}  \label{eq:likeli_2}
& p(d^{\rm GW}_i, d^{\rm EM}_i, D_L, z, \alpha, \delta|\vec{H}, I)   \nonumber \\
= &  ~p(d^{\rm GW}_i, d^{\rm EM}_i | D_L, z, \alpha, \delta, \vec{H}, I) p_0(D_L, z, \alpha, \delta | \vec{H}, I)    \nonumber \\
= &  ~p(d^{\rm GW}_i| D_L, \alpha, \delta, I) 
p(d^{\rm EM}_i| z, \alpha, \delta, I)\nonumber\\
&\times p_0(D_L | z, \vec{H}, I) p_0(z, \alpha, \delta | \vec{H}, I). 
\end{align}
The term $p(d^{\rm GW}_i | D_L, \alpha, \delta, I)$ is the marginalized
likelihood of the \ac{GW} data \citep{Finn:1992wt}, and $p(d^{\rm EM}_i| z,
\alpha, \delta, I)$ is the likelihood of the observational data of the \ac{EM}
counterpart.   
For dark sirens, there is no information for their \ac{EM} counterparts, so we can
set $d^{\rm EM}_i$ to null, as well as set the likelihood
$p(d^{\rm EM}_i| z, \alpha, \delta, I) $
to constant
\citep{Chen:2017rfc}. 
The term $p_0(D_L | z, \vec{H}, I)$ is a Dirac delta function,
i.e., $p_0(D_L | z, \vec{H}, I) = \delta_{\rm D} \big( D_L - \hat D_L(z,
\vec{H}) \big)$, since  $D_L$ is uniquely determined by $z$ given a
cosmological model.  

In Equation~(\ref{eq:likeli_2}),  $p_0(z, \alpha, \delta | \vec{H}, I)$ is a prior.
To calculate it,
we adopt the standard assumption that it is proportional to
the number density of galaxies \citep{Schutz:1986gp, MacLeod:2007jd, Petiteau:2011we, DelPozzo:2011vcw, Chen:2017rfc, LIGOScientific:2018gmd, Gray:2019ksv,
Finke:2021aom}. Taking into account the fact that many galaxies are too dim to be detected,
we write the prior as  
\begin{align} \label{eq:prior_totgalaxy} 
  p_0(z, \alpha, \delta | \vec{H}, I) =& ~ ~\! p_{\rm obs}(z, \alpha, \delta|\vec{H}, I)  \nonumber \\
&  +  ( 1- f_{\rm compl}) ~\! p_{\rm miss}(z, \alpha, \delta | \vec{H}, I),
\end{align}
where $p_{\rm obs}(z, \alpha, \delta | I)$ is a distribution function compiled
from galaxy surveys, $p_{\rm miss}(z, \alpha, \delta | \vec{H}, I)$ represents
the distribution of the undetected galaxies, and $f_{\rm compl}$ is a fraction
denoting the completeness of the galaxy catalog \citep[also
see][]{Chen:2017rfc, LIGOScientific:2018gmd}.

The value of $p_0(z, \alpha, \delta | \vec{H}, I)$ is determined as follows.
(i)
For the observed galaxy distribution function, we calculate it with 
\begin{align} \label{eq:prior_obsgalaxy}
  p_{\rm obs}(z, \alpha, \delta | \vec{H}, I) = & \frac{1}{N_{\rm gal}} \sum_{j=1}^{N_{\rm gal}} \big[ \mathcal{N}(z|\bar z_j, \sigma_{z;j})   \nonumber \\
  &\times \delta_{\rm D} (\alpha - \alpha_j) \delta_{\rm D} (\delta - \delta_j) \big], 
\end{align}
where $N_{\rm gal}$ is the total number of the observed galaxies and
$\mathcal{N}(z|\bar z, \sigma_{z})$ is a Gaussian distribution of $z$ with
an expectation $\bar{z}$ and a standard deviation $\sigma_z$.
Here, we introduced two delta functions because the sky-localization errors
in galaxy surveys are much smaller than those in GW observations.
(ii) For the undetected galaxies, we adopt the expression
%
\begin{align} \label{eq:missing_catalog}
p_{\rm miss}(z, \alpha, \delta | \vec{H}, I) \propto & \big[1 - p_{\rm compl}(z, \alpha, \delta) \big]  \frac{\D^2 V_{\rm c}}{\D z \D \hat\Omega} 
\end{align}
from \citet{Chen:2017rfc}, assuming an  isotropic and homogeneous universe,
where $p_{\rm compl}(z, \alpha, \delta)$ is the probability of a galaxy 
at $(z, \alpha, \delta)$ being in the survey galaxy catalog, 
$V_{\rm c}$ is the comoving volume, and $\D \hat\Omega \equiv \cos \delta ~\D\alpha \D\delta$ represents the differential solid angle. It should be noted here that $\delta$ is defined as the angle 
with respect to the celestial equatorial plane. 
(iii) The completeness fraction is calculated with 
%
\begin{align} \label{eq:completeness_fraction}
f_{\rm compl} = \frac{1}{V_{\rm c}^{\rm tot}} \! & \int \!\!\! \! \int_{4\pi} \!\! \int_{z_{\rm min}}^{z_{\rm max}} \!\!\!   p_{\rm compl}(z, \alpha, \delta)  \frac{\D^2 V_{\rm c}}{\D z \D \hat\Omega} \D z \D \hat\Omega, 
\end{align}
where $z_{\min}(z_{\max})$ is the minimum(maximum) redshift of all the
candidates of the host galaxy of a \ac{GW} source, and $V_{\rm c}^{\rm tot}$ is
the total comoving volume enclosed within $[z_{\min}, z_{\max}]$. 

To calculate the denominator $\beta(\vec{H} | I)$ in Equation
(\ref{eq:likeli_decompos}), we need to consider the selection effects in
observations because we can
only detect those \ac{GW} or
\ac{EM} sources that exceed certain detection thresholds
\citep{Chen:2017rfc}.  Since the \ac{EM} data $d_i^{\rm EM}$ can be set to null
for dark sirens, the denominator is determined solely by the GW sources.
Following \citet{Chen:2017rfc}, we calculate it with
\begin{align} \label{eq:belta_term}
\beta(\vec{H} | I) =& \int_{d^{\rm GW}_i > {\rm threshold}}  \int \!\!\!\! \int \!\!\!\! \int \! p(d^{\rm GW}_i| \hat D_L(z, \vec{H}), \alpha, \delta | I)   \nonumber \\
& \times p_0^{\rm c}(z, \alpha, \delta | \vec{H}, I) ~\! \D d^{\rm GW}_i \D z  \D \hat\Omega, 
\end{align}
where $p_0^{\rm c}(z, \alpha, \delta | \vec{H}, I)$ is a cosmological prior distribution of all possible galaxies. 
For an isotropic and homogeneous universe, we have $p_0^{\rm c}(z, \alpha, \delta | \vec{H}, I) = p_0^{\rm c}(z| \vec{H},I)p_0^{\rm c}(\alpha, \delta|I)$. 
Moreover, we define
\begin{align} \label{eq:GW_frt}
p^{\rm GW}_{\rm det}(D_L |I ) \equiv & \int_{d^{\rm GW}_i > {\rm threshold}} \int \!\!\!\! \int \! p(d^{\rm GW}_i |D_L, \alpha, \delta, I)   \nonumber \\
& \times p_0^{\rm c}(\alpha, \delta | I) ~\! \D d^{\rm GW}_i \D \hat\Omega,
\end{align}
and Equation~(\ref{eq:belta_term}) then becomes 
\begin{align} \label{eq:belta_term-simplified}
\beta(\vec{H} | I) = \int p^{\rm GW}_{\rm det} \big( \hat D_L(z, \vec{H}) |I \big) p_0^{\rm c}(z | \vec{H}, I) \D z. 
\end{align}
The term $p^{\rm GW}_{\rm det}(D_L |I)$ describes the probability that a
\ac{GW} event at $D_L$ will be detected. The other term $p_0^{\rm c}(z |
I)$ describes the probability that a galaxy is at redshift $z$
and can be calculated with
\begin{align} \label{eq:p0_observed} 
& p_0^{\rm c}(z | \vec{H}, I)   \nonumber \\
 \propto & \frac{1}{2 \Delta z} \! \int_{(z-\Delta z)}^{(z+\Delta
z)} \!\!\! \int \!\!\! \! \int_{4\pi} \!  p_0(z', \alpha, \delta | \vec{H}, I)   \D
\hat\Omega  \D z'
\end{align} 
\citep{Zhu:2021aat, Zhu:2021bpp}, where $\Delta z$ is a smoothing interval
that needs to be much larger than the typical redshift scale of a galaxy
cluster, and the expression of the integrand $p_0(z', \alpha, \delta | \vec{H}, I)$ is shown in Equation (\ref{eq:prior_totgalaxy}).  

\subsection{True and false candidate host galaxies}

The effectiveness of using dark sirens to constrain cosmological parameters
relies on deriving a statistically correct distribution function for the
redshift of a GW source. Two factors could potentially impair the reliability
of the derived redshift.  (i) The galaxy catalog is likely incomplete, such that
it may miss the true host galaxy of a GW source. (ii) If the models of the
\ac{GW} signal or the detector noise are inaccurate, the spatial localization of a GW
source will be biased.

The first factor is not catastrophic. 
Given that most galaxies form in clusters, the
brightest galaxies in a galaxy cluster still provide a statistically correct
redshift even if the true host galaxy of the dark siren is too dim to be
observed.  Although the redshifts of the brightest cluster members and the
redshift of the true host galaxy are not exactly the same, the difference has
been accounted for in our Bayesian framework by the completeness fraction
$f_{\rm compl}$ in Equation (\ref{eq:prior_totgalaxy}).  For simplicity, in the
following, we refer to the galaxies that are physically correlated with the
true host galaxy of a dark siren as the ``true candidate host galaxies''.

The second factor is more detrimental.  In this case, the galaxies 
found at the luminosity distance and sky location inferred from the GW data may be
physically uncorrelated with the GW source.  Therefore, the redshift distribution provided
by the galaxies in the incorrect error volume is inconsistent with the true redshift of the
dark siren. This mismatch could result in a large bias in the inference of the
cosmological parameters. These galaxies are referred to as the ``false candidate host galaxies'' in this work.

\section{Mock observation}    \label{sec:Simulations}

To understand the impact of the false candidate host galaxies on the
estimation of the cosmological parameters, we will
conduct mock observations of the host galaxies of the dark sirens, and simulate
future observational constraint on the Hubble-Lema\^itre constant using such galaxies.
Our simulation is divided into three steps.
(i) Generate a mock catalog of merging \acp{SBBH} and distribute them in the galaxies
selected from a cosmological numerical simulation. 
(ii) Conduct mock observations of the candidate host galaxies of the \acp{SBBH} to derive
the statistical probability distribution of the redshift for each binary.  
(iii) Calculate the total posterior
probability distributions of the cosmological parameters using the
luminosity distances of the dark sirens and the statistical redshifts
provided by the candidate host galaxies. The details are as follows.

First, we generate a population of \acp{SBBH} using the ``Power Law
+ Peak'' model presented in \citet{LIGOScientific:2021psn}. 
The direction of the orbital angular momentum is assumed to be random.
The spin parameter is uniformly distributed in the range of $[-1, 1]$.
The merger time is chosen randomly between $0$ and $5$ years,
and the merger phase is also random.
Given these parameters, the \ac{GW} signal is generated using
the IMRPhenomPv2 waveform model \citep{Hannam:2013oca, Schmidt:2014iyl}. 

For each binary in our mock catalog, we assign a host galaxy that is randomly
selected from the galaxy catalog of the MultiDark cosmological simulations
\citep{Klypin:2014kpa, Croton:2016etl, Conroy:2008mp}.  We note that the
cosmological parameters used in the simulations are $H_0 = 67.8 ~{\rm
km/s/Mpc}$ and $\Omega_M = 0.307$.  Because we are mainly interested in studying
the expansion of the late Universe, we restrict the selected
galaxies to $z < 1$.

To evaluate the detectability of a \ac{SBBH} \ac{GW} event, we consider a network composed
of the \ac{ET} \citep{Punturo:2010zz} and the \ac{CE}
\citep{LIGOScientific:2016wof}.  The models for the sensitivity curves are ET-D
\citep{Hild:2010id} and CE2 \citep{Reitze:2019iox, Reitze:2019dyk}.  The
response of the network to a \ac{GW} signal is computed using the low-frequency
approximation \citep{Thorne:1989lfp}, where we have taken the rotation of the
Earth into account \citep{Zhao:2017cbb}. 

\begin{figure}[htbp]
\centering
\includegraphics[width=0.450\textwidth]{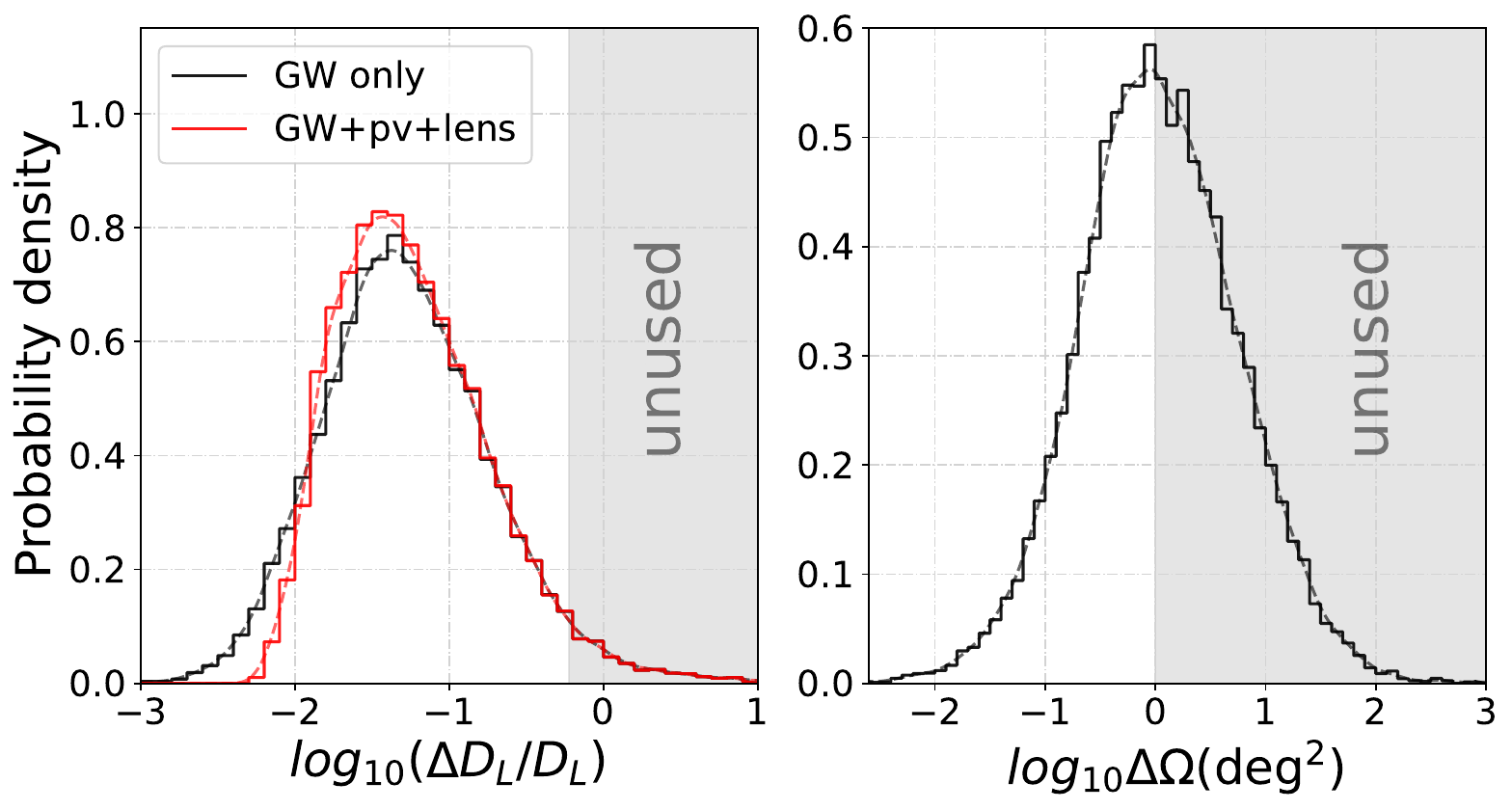}
\caption{Distribution of the spatial localization errors for the \acp{SBBH} detected by the ET+CE network. 
The left panel shows the error of the luminosity distance. The black line represents the error 
caused by the detector noise, and the red line
further accounts for the peculiar velocities of the galaxies and the weak-lensing effect. 
The right panel shows the sky-localization error.   }
\label{fig:SBBH_dDLdOmega_errs}
\end{figure}

Second, for each detectable \ac{SBBH}, we select the \emph{candidate} host galaxies from
the same MultiDark galaxy catalog according to the spatial localization
inferred from the \ac{GW} data. These galaxies are used to compile the
probability distribution of the redshift of the dark siren.  There are multiple candidates because there are large errors in the inferred luminosity distance and
sky location.  In our simulation, these errors are derived according to the
Fisher information matrix \citep{Cutler:1994ys, Poisson:1995ef,
Vallisneri:2007ev} with a $3\sigma$ \ac{CI}. In particular, we have also
included additional errors in luminosity distance caused by the peculiar velocities of the galaxies and the
weak-lensing effect \citep{Kocsis:2005vv, Hirata:2010ba}.  Figure
\ref{fig:SBBH_dDLdOmega_errs} shows the spatial localization errors, including
the error in luminosity distance, $\Delta D_L/D_L$, and the error in sky
localization $\Delta \Omega$.  We can see that, on average, $\Delta
D_L/D_L\simeq4\%$ and $\Delta \Omega\simeq1\,{\rm deg}^2$. In this study, we
only use those \acp{SBBH} with relatively small localization errors, namely $\Delta
D_L /D_L \leq 0.6$ (equivalent to $\Delta D_L/D_L \lesssim 1$ at the confidence level
of $90\%$) and $\Delta \Omega \leq 1 ~ {\rm deg}^2$, because the sources with
larger localization errors have less constraining power on the cosmological
parameters.

%

It is important to mention that luminosity distances are not directly observable
in galaxy surveys. The real observables are the redshifts of galaxies.
Therefore, we must convert the luminosity distance inferred from
GW data into
the information on redshift before we can select the candidate host galaxies
for a dark siren.
The conversion is performed using the relationship 
\begin{subequations}
 \begin{align}
 D_{L;\min} &=  c(1+z_{\min}) \int_{0}^{z_{\min}} \frac{1}{H^-(z')} \D z',    \label{eq:DL_min}    \\
 D_{L;\max} &=  c(1+z_{\max}) \int_{0}^{z_{\max}} \frac{1}{H^+(z')} \D z',    \label{eq:DL_max}
 \end{align}
\end{subequations}
where $H^{-}(z)$ and $H^+(z)$ are the minimum and maximum values of $H(z)$
given by the prior of the cosmological parameters, $[D_{L;\min}, D_{L;\max}]$
is the range of luminosity distance allowed by the GW data, and $z_{\min}$ and
$z_{\max}$ are the resulting minimum and maximum values for the redshift.  For
the prior, we choose a uniform distribution of $H_0$ in the range of $[30, 120]
~{\rm km/s/Mpc}$ and a uniform distribution of $\Omega_M$ in $[0.04, 0.6]$.  We
do not consider the observational errors in the spectroscopic redshifts of the
host galaxies, because they are small in the redshift range of our interest
\citep{DESI:2016fyo, Gardner:2006ky, Gong:2019yxt}.  Given such redshift
information, the candidate host galaxies of a dark siren are selected from an
``error volume'' of a size of $9 \Delta \Omega \times [z_{\min}, z_{\max}]$ in the
MultiDark galaxy catalog.

Moreover, we must include the selection effect that exists in real galaxy surveys. 
In our simulation, the probability that a galaxy with an apparent magnitude of 
$\bar m^*$ is observed is
\begin{equation} \label{eq:selection_function}
{\rm erfc} (\bar m^*) = \frac{1}{\sqrt{2\pi } \sigma_{m^*}} \! \int^{m^*_{\rm limit}}_{-\infty} \!\! \exp \! \bigg( \!\! - \! \frac{1}{2} \frac{(m^* - \bar m^*)^2}{\sigma_{m^*}^2} \bigg) \D m^* \!,
\end{equation}
where $\sigma_{m^*}$ is the uncertainty in measuring the magnitude and
$m^*_{\rm limit}$ is the limiting magnitude of the galaxy survey. 
We assume a conservative value of $\sigma_{m^*} = 0.1 ~{\rm mag}$
and use $m^*_{\rm limit} = +26 ~{\rm mag}$ to mimic future observations by the
Chinese Space Station Telescope (CSST) \citep{Gong:2019yxt}.

By now, we have selected the true candidate host galaxies. However, we also need to
simulate the observations of the false candidate host galaxies in order to understand their
potential impact.  Previous studies have shown that the dark sirens at
different cosmological distances will lead to systematically different constraints
on the cosmological parameters \citep{Zhu:2021aat}.  To eliminate this
statistical difference correlated with distance and single out the effects
caused by the incorrect identification of the candidate host galaxies, we simplify our
problem by keeping the luminosity distance and its error constant, while changing only the sky localization.
Therefore, we introduce a displacement to the true sky localization. 
Using this displaced sky localization and the same
luminosity distance information, we select the false candidate host galaxies from the incorrect error volume. 
In this way, the selected galaxies are physically uncorrelated with
the true host galaxy of the dark sirens, but in the redshift space the true
and false candidate host galaxies share the same distribution. 
In this work, the values of the displacement,  $(\D \alpha, \D \delta)$,
are randomly generated within the boundary of the sky area of the galaxy survey. 
To ensure that the displaced sky position does not 
overlap with the true position of the \ac{GW} source,
we also impose a lower limit to the displacement, which is determined by
$(\D \alpha, \D \delta) \Sigma^{-1}_{\alpha \delta} (\D \alpha, \D \delta)^{\rm T}/2 > 11.83$ 
where $\Sigma_{\alpha \delta}$ is the covariance matrix of the 
sky localization of the dark siren. This way,
the inconsistency between the displaced and the true sky localization is not less than $3 \sigma$.

Third, we use the statistical redshift distributions of true and false candidate
host galaxies to compute the posterior probability distributions of the
cosmological parameters. The calculation is based on Equations (\ref{eq:posterior}).
In the calculation, we use the \ac{MCMC} method to accelerate the calculation. 
Moreover, we use the \textsf{emcee} library, which is a Python-based affine invariant 
\ac{MCMC} ensemble \citep{ForemanMackey:2012ig, ForemanMackey:2019ig}.

\section{Results}    \label{sec:ResultsAnalyses}

\subsection{Constraints on the cosmological parameters}

One merit of using dark sirens to constrain cosmological parameters is that the
precision improves with the number of dark sirens. This may no
longer be true when the dark sirens with false candidate host galaxies are present. Therefore, we first study
the precisions of the cosmological parameters as a function of the number of
dark sirens, $N$. 

We consider a number of values for $N$, namely $N=10$, $30$, $100$, $300$, and
$1000$, which are almost equally spanned in the logarithmic space. Our maximum
number $N=10^3$ is reasonable because (i) the ``Power Law+Peak'' population
model \citep{LIGOScientific:2021psn} for \acp{SBBH} predicts that about
$16,000-35,000$ \acp{SBBH} at $z \leq 1$ would merge within a period of $5$
years, and (ii) about one-third of them could be covered by the current and
future galaxy surveys \citep{DESI:2016fyo, Gong:2019yxt, Euclid:2021icp}.

\begin{figure}[htbp]
\centering
\includegraphics[width=0.450\textwidth]{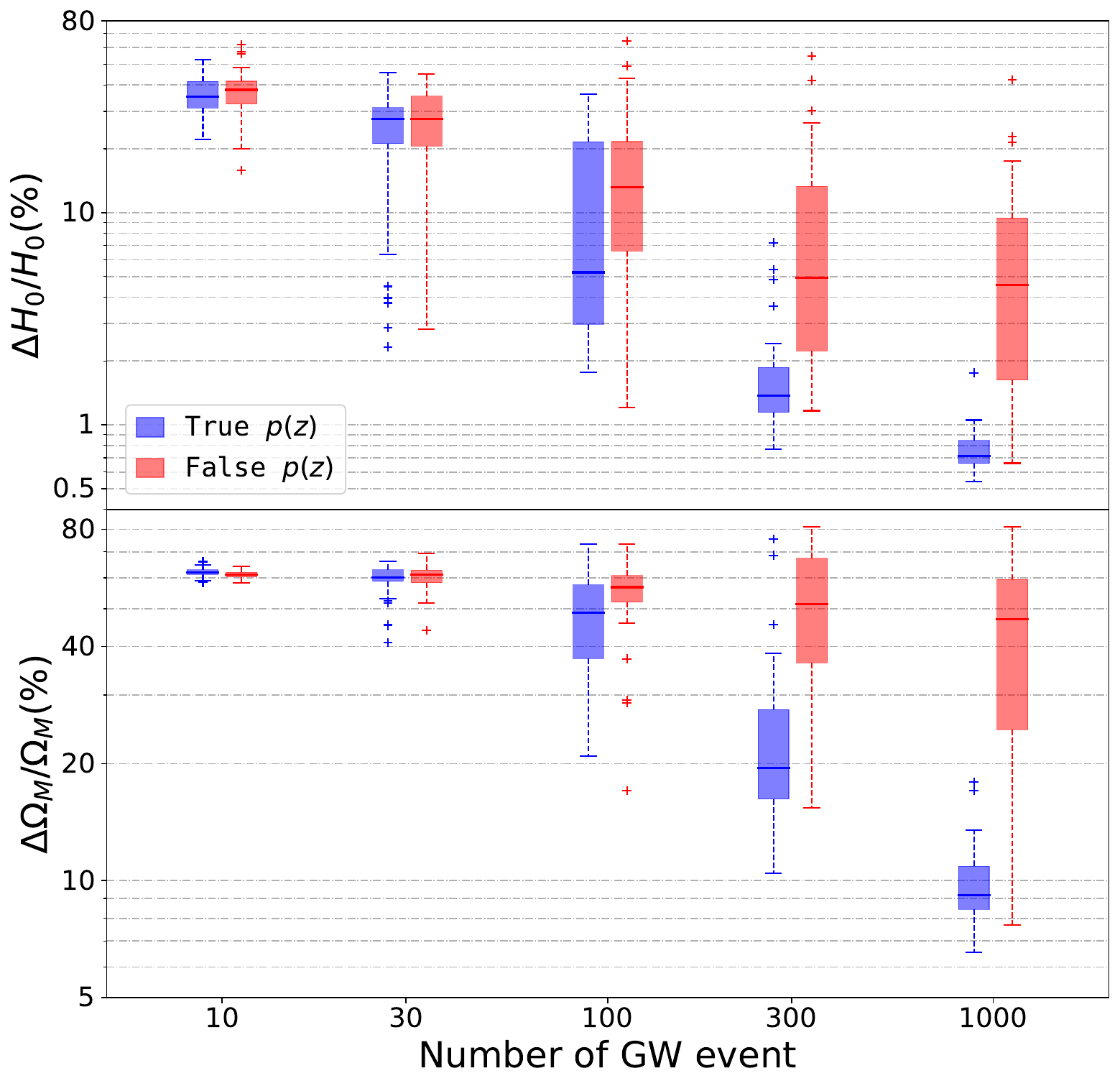}
\caption{Dependence of the precisions of $H_0$ (upper panel) and $\Omega_M$ (lower panel) 
on the number of dark sirens. The blue and red colors refer to the results based on,
respectively, true and false candidate host galaxies.
The boxes show the  $25\%-75\%$ \ac{CI} derived from 48 independent trial simulations.
The short horizontal line inside each box shows the median value, and the whiskers mark
a maximum length that is
$1.5$ times the length of the box (approximately 99\% \ac{CI} if the data distribution is Gaussian).
The crosses are the outliers of the trials. 
}
\label{fig:TrueFalse_H0_RelativeErrs}
\end{figure}

Figure~\ref{fig:TrueFalse_H0_RelativeErrs} shows the dependence of the relative
errors of $H_0$ and $\Omega_M$ on $N$.  The median values for the errors are
shown as the short horizontal bars inside the boxes, while the boxes show the
$25\%-75\%$ \ac{CI} based on 48 trials.  When $N\le100$, the error
corresponding to true candidate host galaxies is statistically comparable with
the error derived from false candidate host galaxies.  However, as soon as $N$
reaches $300$, the difference between the two errors becomes statistically
significant: the latter becomes substantially larger than the former.  In
particular, when $N=10^3$ the median precisions of $H_0$ and
$\Omega_M$ become better than $1\%$ and $10\%$, respectively, if true candidate host
galaxies are used in the simulations.  On the other hand, the median precisions
corresponding to false candidate host galaxies are worse by a factor of about five.

\begin{figure}[htbp]
\centering
\includegraphics[width=0.450\textwidth]{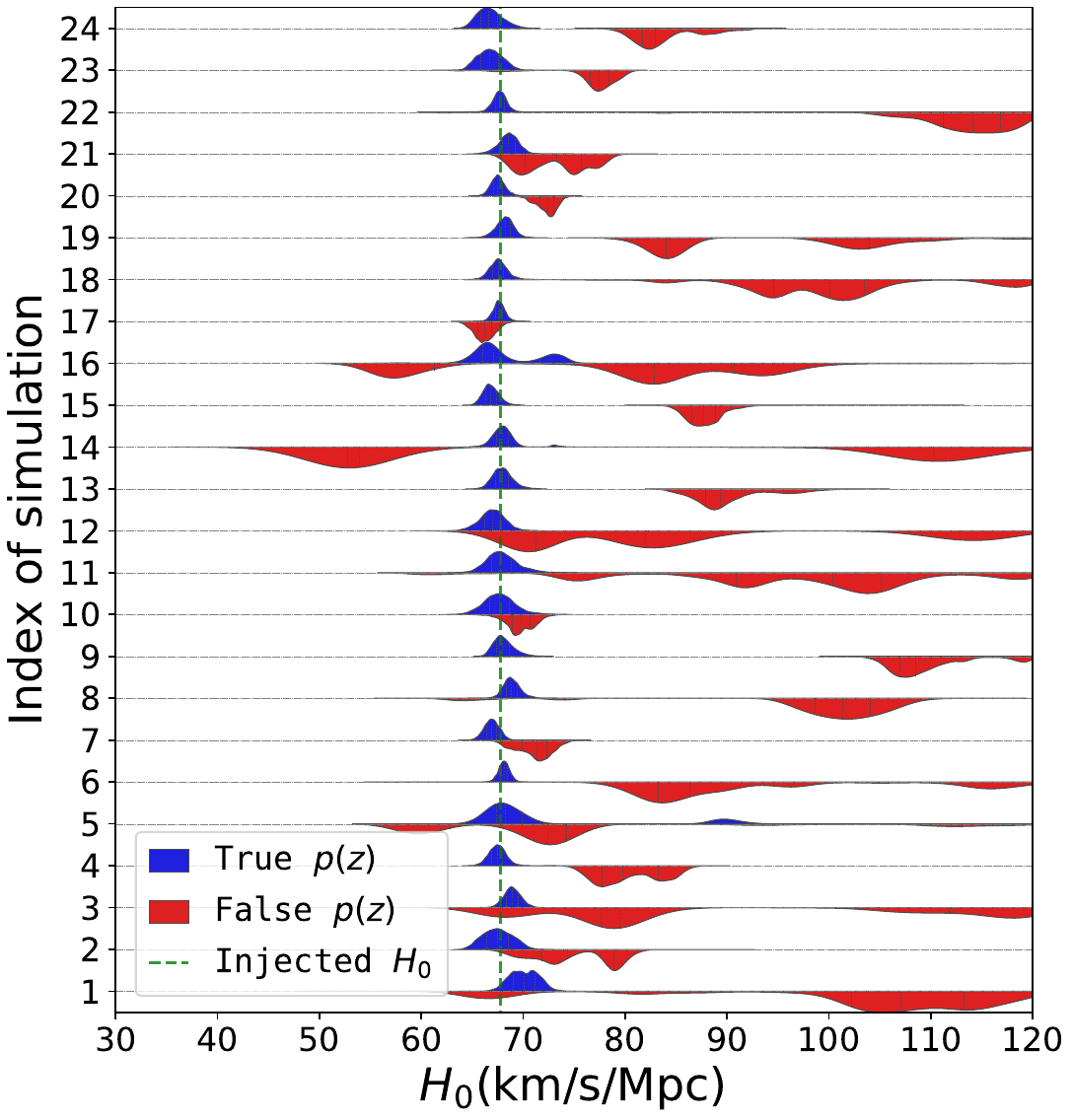}
\caption{Posterior distribution of $H_0$ in $24$ independent trial simulations,
each simulation containing $300$ dark sirens. The upper blue (lower red) distributions
correspond to the simulations using the true (false) host galaxies.
The vertical green dashed line marks the correct value of $H_0$.}
\label{fig:TrueFalse_H0_Violinplot}
\end{figure}

Figure~\ref{fig:TrueFalse_H0_RelativeErrs} also shows that, as soon as $N$
reaches $300$, the spread of the $25\%-75\%$ \ac{CI} (the length of a box)
becomes statistically different in the two kinds of simulations. 
The spread is systematically wider in the simulations
with false candidate host galaxies. 
The reason can be found in
Figure~\ref{fig:TrueFalse_H0_Violinplot}, where we show the posterior
distributions of $H_0$ from the first 24 trial simulations when $N=300$.  
We find that, when we use true candidate host
galaxies (blue), the posterior distribution in each trial simulation is
relatively concentrated around the true value of $H_0$.  Moreover, only in
three trials (1, 5, and 16) do we see significant secondary peaks offset from the
true value. On the other hand, the simulations using false candidate host galaxies
(red) more often produce multiple peaks, and the locations of the peaks are
widespread. Such randomness is a direct consequence of the incorrect redshift
distribution compiled from the incorrect candidate host galaxies. 

We notice that, when $N = 100$, the \acp{CI} given by the
simulations with true candidate host galaxies are much broader than those given by
false candidate host galaxies.  This is because, when the number of dark sirens
is small, e.g.  $N=100$, the \ac{GW} catalog  may or may not contain very
precisely localized dark sirens (e.g., $\Delta \Omega \sim 0.01 {\rm deg}^2$).
Without such well-localized dark sirens, the constraint on cosmological
parameters would be relatively weak.

The above results suggest that, in the future, with about $300$ dark sirens, we
can start to assess the reliability of our modeling of the \ac{GW} signals.  
A small dispersion in the posterior distribution of the cosmological parameters
may suggest that the selected galaxies are indeed physically correlated with the true host galaxies of the dark sirens. 
Therefore, the model which we use to infer the spatial localization of a \ac{GW} source is
accurate. If the posterior distribution has a large dispersion, the model is
unlikely to be accurate, and it may miss the effects caused by the astrophysical
environments, as we have discussed in Section~\ref{sec:Introduction}.  

\subsection{Testing the reliability of the candidate host galaxies}

Besides evaluating the posterior distribution functions of the cosmological parameters,
we also design two other methods that can be used to infer the reliability of
the selection of the candidate host galaxies for dark sirens.

The first method is a translation of the sky position. In this method, we
search for the candidate host galaxies of a dark siren in two different error volumes: an
original error volume derived directly from the \ac{GW} data, and the other is obtained by
artificially shifting the sky position of the original error volume by a large
angle. We then compute the posterior distribution function of $H_0$
using the galaxies selected from the above two error volumes and derive 
the corresponding error $\Delta H_0$.  The reliability
of the host galaxies can be inferred as follows.  (i) If the original error volume
contains the true host galaxy, the error $\Delta H_0$ should increase significantly after
the translation of the sky position, because the shifted error volume contains
only physically unrelated false candidate host galaxies.  (ii) If, on the other hand, the
original error volume contains only false candidate host galaxies, the $\Delta H_0$ before
and after the translation of sky position should be statistically comparable.

\begin{figure}[htbp]
\centering
\includegraphics[width=0.450\textwidth, height=0.290\textwidth]{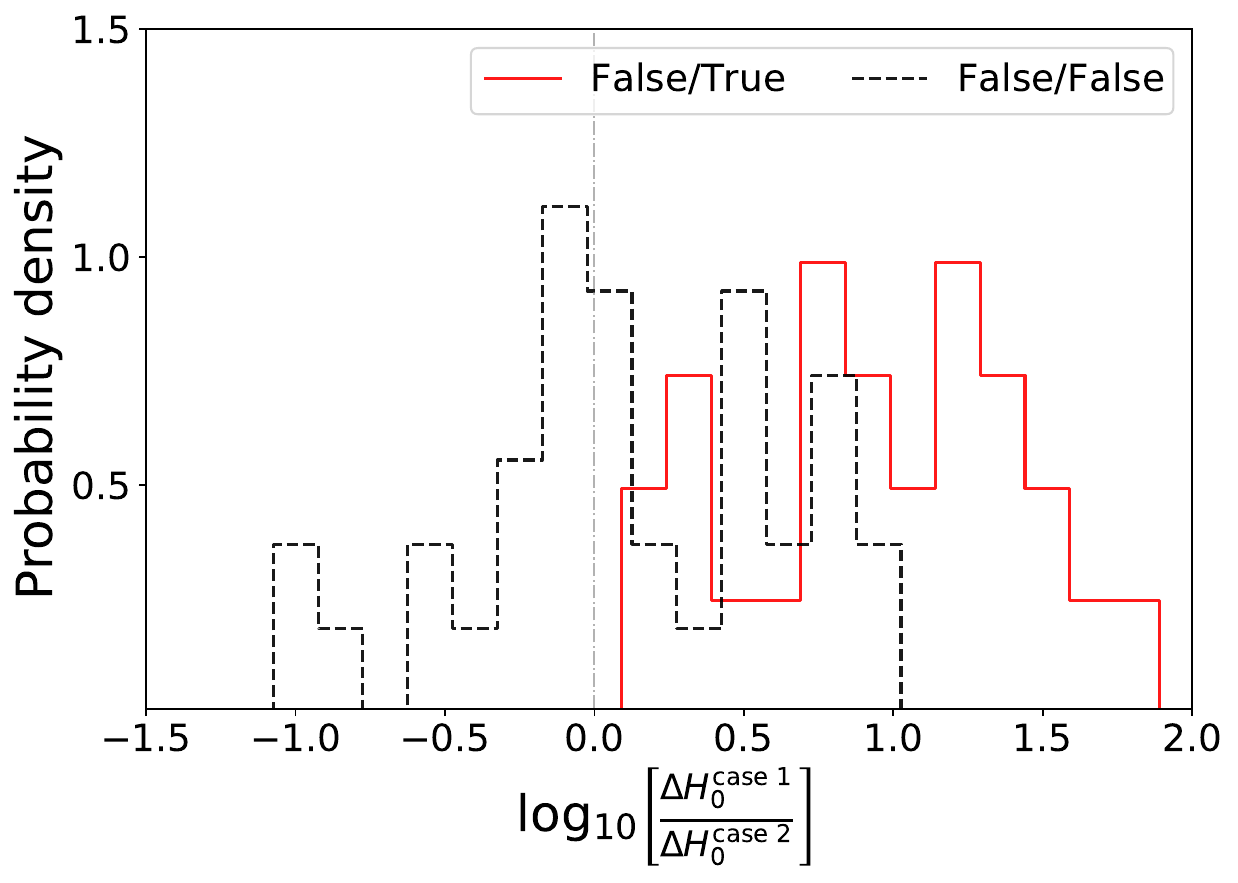}
\caption{Probability of the ratio between two errors of $H_0$,
one ($\Delta H_0^{\rm case\,2}$) derived using the galaxies in the original 
error volume inferred from \ac{GW} signal and the other ($\Delta H_0^{\rm case\,1}$) using the galaxies
in an error volume shifted from the original one. Each error is compiled from
$300$ dark sirens.
The red solid histogram corresponds to the scenario in which 
the original error volume contains true candidate host galaxies, and
the black dashed histogram corresponds to the scenario in which the 
original error volume only contains false candidate host galaxies.
}
\label{fig:H0_RelativeErrs_Ratio}
\end{figure}

Figure~\ref{fig:H0_RelativeErrs_Ratio} shows the probability distribution of the ratio
between the two errors, one $\Delta H_0$ derived before the sky translation and
the other after the sky translation. If the original error volumes (before the
sky translation) contain the true host galaxy (red solid histogram), we find
that, in all our simulations, the ratio will be greater than unity, i.e., the
error $\Delta H_0$ worsens after the sky translation.  In particular, the
ratio may exceed $10$ in about $43\%$ of the simulations.  If the original
error volume contains only false candidate host galaxies (black dashed histogram),  the
probability that the ratio of $\Delta H_0$ is greater than unity is comparable
to the probability of being smaller than unity. Moreover, the ratio hardly
exceeds $10$ in this latter case. 
These results are generally consistent with our earlier
prediction that a sky translation worsens the constraint on cosmological parameters
if the original sky localization is accurate. On the contrary, the sky translation
does not significantly affect the constraint if the original sky localization
is inaccurate.

Based on the above results, we suggest a value of $10$ as the threshold
of verifying the reliability of the sky localization of dark sirens. 
If the error of $H_0$ increases by a factor of $10$ after a sky translation,
the original error volumes where we search for the host galaxies of dark sirens
are likely to be correct. If the error of $H_0$ increases by a factor less than
$10$, the original error volumes should be taken with caution because they may
miss the host galaxies in about 57\% of the cases.

We note that our method is based on ideal assumptions about the completeness of the
galaxy catalog.  In practice, it is important to ensure that the survey
depth and dust extinction are comparable in the error volumes before and after
the sky translation. Otherwise, the completeness of the galaxy catalog will
also affect the errors of the cosmological parameters.

The second method we use to test the reliability of the candidate host galaxies is
the consistency check \citep{Petiteau:2011we}. It divides dark sirens into
smaller groups and performs Bayesian inference independently on each
group. If the spatial localization is correct and the candidate host galaxies are
reliable, the posterior probabilities for different groups should be
consistent. Alternatively, if the spatial localization is mostly incorrect and the
selected host galaxies are false, the posteriors will be more random.

\begin{figure}[htbp]
\centering
\includegraphics[width=0.450\textwidth]{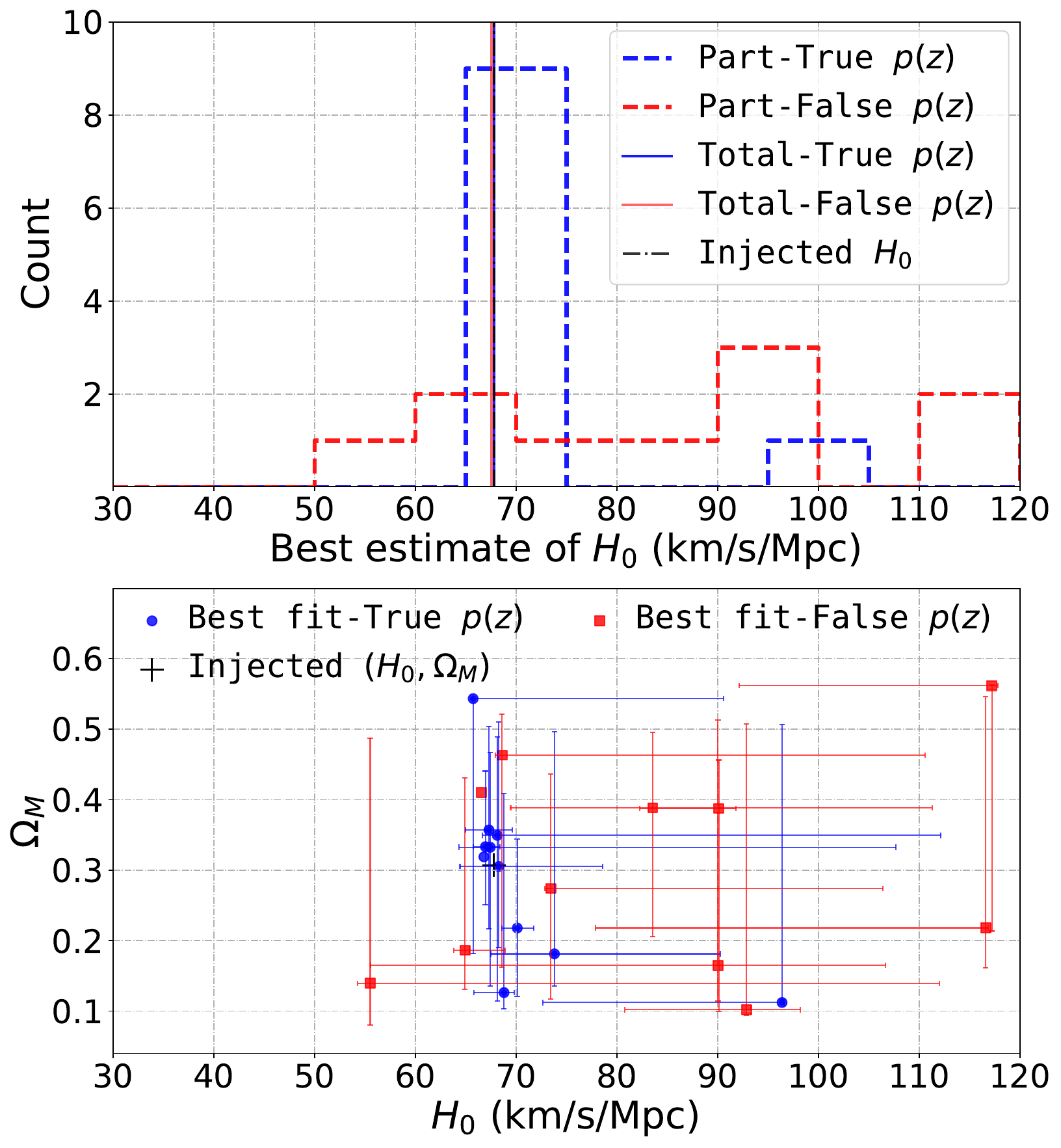}
\caption{Consistency check using a total number of $1000$ dark sirens which are dived
into $10$ nonoverlapping groups.
Upper panel: Distribution of the best estimates of $H_0$ (histograms). The vertical solid lines
 mark the best-estimate values derived from all the  $1000$ dark sirens.
Lower panel: Scatter plot of the estimation results of $(H_0, \Omega_M)$.
The error bars represent the $68.27\%$ \ac{CI}. 
In both panels, the blue (red) color corresponds to the results derived from ture (false)
host galaxies.
}
\label{fig:ConsistencyCheck}
\end{figure}

Figure~\ref{fig:ConsistencyCheck} shows the result of such a test using $1000$
dark sirens. Without dividing into groups and analyzing as a whole population, 
the $1000$ dark sirens will result in a similar posterior distribution 
(see the vertical lines in the upper panel) regardless of
the kind of host galaxies we use (as see Appendix \ref{appendix:TrueFalse_similar_corner}).  However, if we divide them into $10$
groups and perform Bayesian inference separately, the results for true and
false candidate host galaxies become different. We can see that the posteriors of the
$10$ groups are consistent when the true candidate host galaxies are used in the
simulation (blue), but the posteriors largely differ when we use false candidate host
galaxies (red).

\section{Discussion}    \label{sec:Discussions}

So far, we have studied the cases in which the fraction of false candidate host galaxies
for dark sirens is either $f=0$ or $f=100\%$. In reality, both true and false candidate host
galaxies may exist in the sample.  A higher $f$
normally results in a larger bias in the estimation of the cosmological parameters.
Here, we use a percentile-percentile (P-P) plot \citep{doi:10.1198/106186006X136976} to demonstrate
the dependence of the bias on the fraction of the false candidate host galaxies.

\begin{figure}[htbp]
\centering
\includegraphics[width=0.450\textwidth]{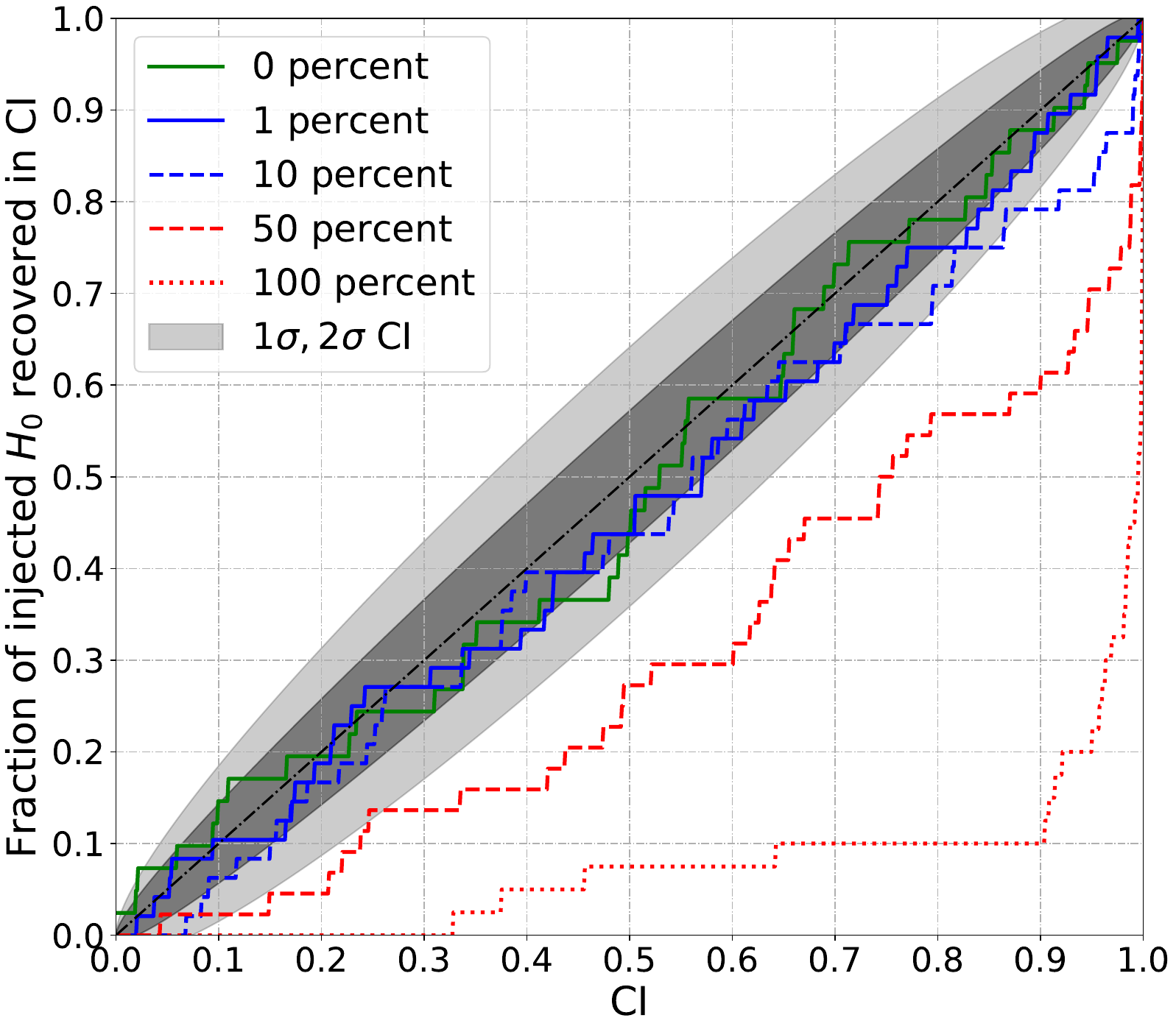}
\caption{Fraction of trial simulations with injected $H_0$ within a \ac{CI} as a function of \acp{CI}. 
The histograms of different colors show the results for different fractions of false candidate host galaxies.
Each histogram is compiled from $48$ independent simulations, and each simulation
contains $300$ dark sirens. 
The dark and light shaded areas show, respectively, the $1 \sigma$ and $2 \sigma$ \acp{CI} allowed for the statistical errors. 
We perform the Kolmogorov-Smirnov (KS) test, with a null hypothesis that
the histogram is consistent with the diagonal. The $p$-values are 0.61, 0.79, 0.28, $2.4 \times
10^{-4}$ and $1.8 \times 10^{-28}$, respectively, when $f= 0\%, 1\%, 10\%, 50\%$, and $100\%$.
}
\label{fig:pp-plot}
\end{figure}

An example is shown in Figure~\ref{fig:pp-plot}. 
When all the spatial localizations are accurate, 
the estimation of $H_0$ is
unbiased, 
the fraction of trial simulations whose true values of $H_0$ fall within specific \acp{CI} as a function of the corresponding \acp{CI} (this is referred to as a P-P plot), should be along the diagonal 
line, as shown by the black dot-dashed line. Statistical fluctuation 
will blur the distribution, as is illustrated by the dark and light shaded areas in the plot, but the result should be still consistent with the diagonal.

When the spatial localizations for a fraction of dark sirens are false, the P-P plot will deviate from the diagonal, as is shown by the colored histograms in
Figure~\ref{fig:pp-plot}. Here, we have used $48$ trial simulations to compile
each histogram, and in each simulation we consider $300$ dark sirens. 
We find that when the fraction of false spatial localization is no more than about $10 \%$, the corresponding
histograms are consistent with the diagonal line within the $2 \sigma$ \ac{CI}.
In this case, the estimation of $H_0$ is more or less unbiased. However, when $f \ge 10\%$, the histograms start to exceed the
$2\sigma$ \ac{CI}, indicating a significant bias in the estimation of $H_0$.
In particular, when $f=50\%$ or $100\%$, the histogram is mostly outside the
$2\sigma$ \ac{CI}. 
This result is not surprising, because now false candidate host
galaxies predominate.

\begin{figure}[htbp]
\centering
\includegraphics[width=0.450\textwidth]{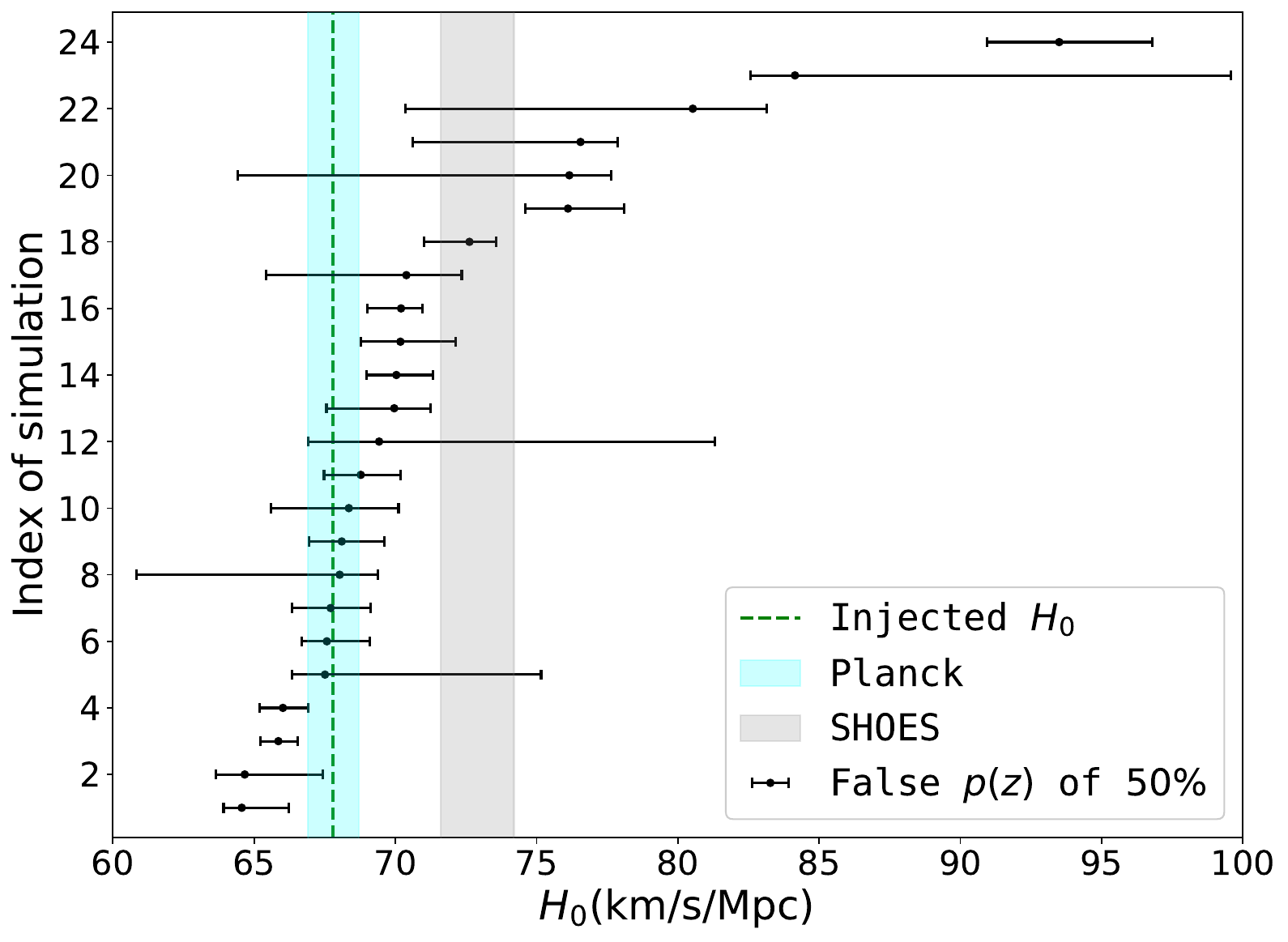}
\caption{The constraint on $H_0$ when half of the dark sirens are localized incorrectly, i.e.,  
$f=50\%$. The results shown here are derived from $24$ independent simulations, each simulation
containing $300$ dark sirens. The error bars show the $68.27\%$ \ac{CI}.
The vertical dashed green line marks the true value of $H_0$ that we injected into the simulation. 
The cyan and gray vertical stripes are the results from the observations of, respectively, 
the \ac{CMB} \citep[Planck,][]{Planck:2015fie} and type-Ia supernovae \citep[SH0ES,][]{Riess:2021jrx}.
	\label{fig:50percentFalse_errbar}}
\end{figure}

Figure~\ref{fig:50percentFalse_errbar} shows more clearly the bias of the estimated
$H_0$ when half of the spatial localizations are false ($f=50\%$).  
We find that the posterior distributions no longer center around the 
injected value of $H_0$. 
In particular, $12$ out of the $24$ simulations show a $> 1 \sigma$
bias between the posterior and the injected  $H_0$.
In seven cases (No. 18-24), we find that the posterior distribution
functions for $H_0$ are more consistent with the observations of type-Ia supernovae even
though the value we injected is clearly from the measurement of \ac{CMB}.
This result highlights the risk of using dark sirens to constrain cosmological parameters
when the environmental effects on \ac{GW} signals are not well understood.

\section{Conclusion}    \label{sec:Conclusions}

Recent studies have shown that the astrophysical environments of \ac{GW} sources
could significantly perturb the \ac{GW} signals. Neglecting such an effect
would lead to systematic biases in the estimation of the distance and sky
position of a \ac{GW} source.  In this work, we have studied the consequences of
such a bias on the measurement of the cosmological parameters such as the
Hubble-Lema\^itre constant.  

Our main findings are as follows. (i) The biased spatial localization will
result in an incorrect statistical probability distribution for the redshift of
a dark siren, which will in turn weaken and bias the constraint on the
cosmological parameters. (ii) The reliability of the statistical redshifts of
dark sirens can be tested by two methods, namely a sky translation and a
consistency check. (iii) The constraints on cosmological parameters by dark
sirens will be significantly biased if more than $10\%$ of the dark sirens are
localized incorrectly.

The above findings are based on the assumption that more than $300$ dark
sirens can be detected and localized well.  It is worth noting that, in our
analysis, we only use the \ac{GW} sources with a localization error of $\Delta
\Omega < 1 ~{\rm deg}^2$. The total number of detected dark sirens is
approximately twice the number of the above well-localized ones.  If the number
of well-localized dark sirens is significantly below $300$, we find that the
presence of false candidate host galaxies does not significantly affect the distribution of $H_0$ error.  The same is true even in the extreme scenario
that all the candidate host galaxies are inferred incorrectly.  This result
indicates that, when the number of dark sirens is low (e.g. $\la100$), the distribution of 
precisions of $H_0$ is limited mainly by the intrinsic uncertainty of the
method, i.e., the statistical uncertainty in compiling the redshift
distribution of the host galaxy of a dark siren.

Several recent studies have predicted that the third generation of ground-based
\ac{GW} detectors can constrain the Hubble-Lema\^itre constant to an precision
as high as $\Delta H_0 / H_0 \lesssim 10^{-4}$ using dark sirens
\citep{Yu:2020vyy, Song:2022siz}.  However, the sky localizations of a few to a
dozen percent of the dark sirens may be biased by nearby SMBHs
\citep[e.g.]{Peng:2021vzr}, strong gravitational lensing
\citep{Broadhurst:2018saj, Broadhurst:2022tjm, Yang:2021viz, Smith:2017mqu}, or
non-stationary detector noise \citep{Edy:2021par, Kumar:2022tto}.  Without
properly modeling these effects, the constraint on the Hubble-Lema\^itre
constant is likely to be much weaker.

\section*{acknowledgments}

This work is supported by the National Natural Science Foundation of China
	grants Nos. 11991053 and 11873022.  We acknowledge the use of the {\it
	TianQin} cluster, a supercomputer owned by the TianQin Research Center
	for Gravitational Physics, Sun Yat-sen University, and of the {\it
	Tianhe-2}, a supercomputer owned by the National Supercomputing Center
	in GuangZhou.  The authors also thank Yi-Ming Hu, Jian-dong Zhang,
	Shuai Liu, and En-Kun Li for helpful discussions, and especially
	Yi-Ming Hu for important suggestions on the structure of this paper. 

\software{
\textsf{numpy} \citep{vanderWalt:2011bqk}, \textsf{scipy} \citep{Virtanen:2019joe}, \textsf{LALSuite} \citep{lalsuite}, \textsf{emcee} \citep{ForemanMackey:2012ig, ForemanMackey:2019ig}, \textsf{matplotlib} \citep{Hunter:2007ouj}, and \textsf{corner} \citep{corner}. 
}


\appendix
\renewcommand\thefigure{\Alph{section}\arabic{figure}}
\renewcommand\thetable{\Alph{section}\arabic{table}}

\section{Similar constraints resulting from different galaxies}    \label{appendix:TrueFalse_similar_corner}

\begin{figure}[htbp]
\centering
\includegraphics[width=0.450\textwidth]{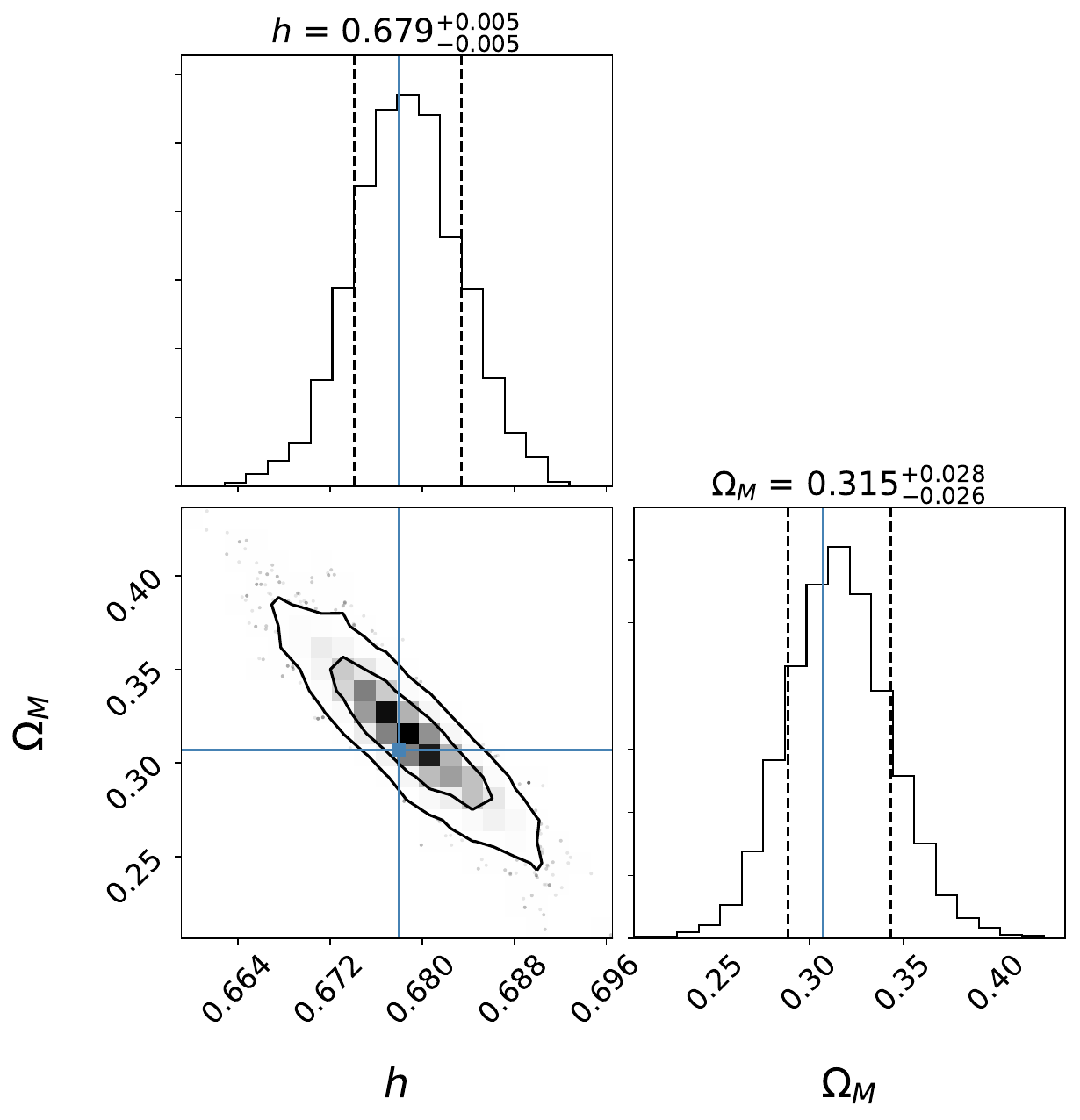} ~~~~~
\includegraphics[width=0.450\textwidth]{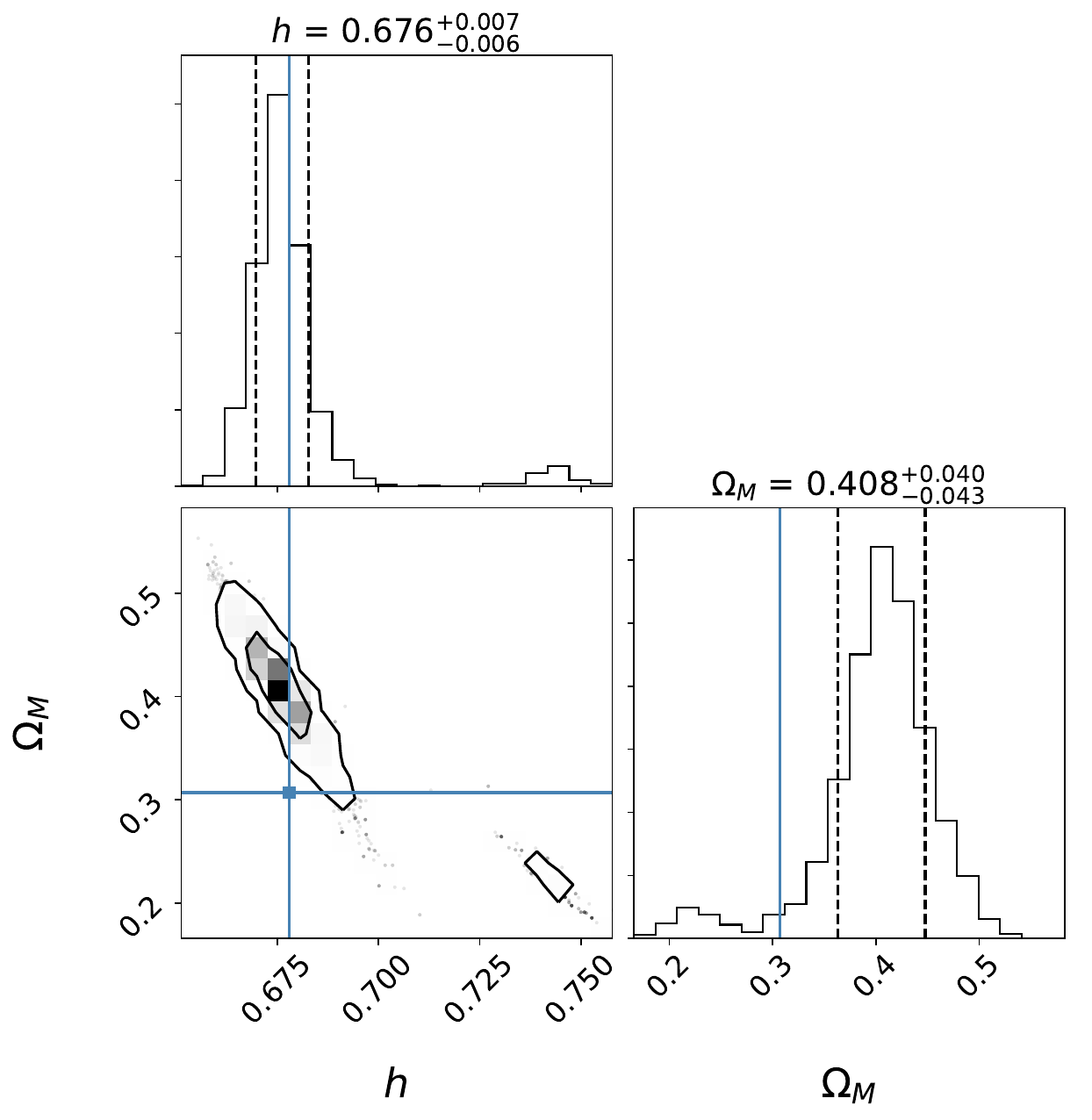}
\caption{
Corner plots of the posteriors for the parameters $h$ ($h \equiv \frac{H_0}{100 ~{\rm km/s/Mpc}}$) and $\Omega_M$ 
derived from true (left) and false (right) candidate host galaxies
of $1000$ dark sirens.
In each subplot, the lower left panel shows the two-dimensional joint posterior 
probability of $h-\Omega_M$, where the contours represent the \acp{CI} of $1 \sigma (68.27\%)$ and $2 \sigma (95.45\%)$. 
The upper and right panels show the one-dimensional posterior probabilities 
for $h$ and $\Omega_M$, which are marginalized over the other parameters.
The dashed vertical lines indicate the $1 \sigma$ \ac{CI}, and  
in each panel the solid cyan lines mark the true values of the corresponding parameters. 
}
\label{fig:cornerPlot_comparison1} 
\end{figure}

Our simulations show that sometimes false candidate host galaxies will result in a
posterior distribution of cosmological parameters that is similar to the
one given by the true candidate host galaxies.  Figure~\ref{fig:cornerPlot_comparison1}
shows an example where we compare the corner plots made from the true (left
subplot) and false (right subplot) host galaxies.  A consistency check based on
this example is presented in Figure~\ref{fig:ConsistencyCheck}.

\bibliography{reference_SBBH_TF}

\end{document}